\documentclass[12pt]{article}
\usepackage{color,graphicx,ifpdf,latexsym,url}
\title{Worst-Case Optimal Tree Layout in External Memory}

\newcommand{\bullshit}[1]{}

\author{%
  Erik D.~Demaine\thanks{%
    MIT Computer Science and Artificial Intelligence Laboratory,
    32 Vassar Street, Cambridge, MA 02139, USA.
    \protect\url{edemaine@mit.edu}.
 Supported in part by NSF grants CCF-0430849,  OISE-0334653 and MADALGO---Center for Massive Data Algorithmics, a Center of the Danish National Research Foundation.}
\and
  John Iacono\thanks{%
    Polytechnic Institute of New York University (Formerly Polytechnic University),
    5 MetroTech Center, Brooklyn, NY 11201, USA and MADALGO---Center for Massive Data Algorithmics, a Center of the Danish National Research Foundation, Aarhus University, \AA bogade 34, 8200 Aarhus N, Denmark.
    Research partially supported by NSF grants CCF-0430849, OISE-0334653, CCF-1018370, and an Alfred P.~Sloan fellowship.     
    \protect\url{http://john.poly.edu}.
  }
\and
  Stefan Langerman\thanks{%
    Directeur de recherches du F.R.S.-FNRS,
    Universit\'e Libre de Bruxelles,
    D\'epartement d'informatique, 
    ULB CP212, Belgium.
    \protect\url{Stefan.Langerman@ulb.ac.be}.
  }
}

\date{}

\advance \topmargin by -\headheight
\advance \topmargin by -\headsep
\evensidemargin \oddsidemargin
\let\latexcite=\cite
\def\cite{\nolinebreak\latexcite}
\let\latexref=\ref
\def\ref{\nolinebreak\latexref}

{\makeatletter
 \gdef\xxxmark{%
   \expandafter\ifx\csname @mpargs\endcsname\relax 
     \expandafter\ifx\csname @captype\endcsname\relax 
       \marginpar{xxx}
     \else
       xxx 
     \fi
   \else
     xxx 
   \fi}
 \gdef\xxx{\@ifnextchar[\xxx@lab\xxx@nolab}
 \long\gdef\xxx@lab[#1]#2{{\bf [\xxxmark #2 ---{\sc #1}]}}
 \long\gdef\xxx@nolab#1{{\bf [\xxxmark #1]}}
}

{\makeatletter \gdef\fps@figure{!htbp}}

\newtheorem{theorem}{Theorem}

\def\f{\mathop{\rm jeffe}\nolimits}
\def\f{\mathop{\rm cost}\nolimits}
\def\LEFT{\ell}
\def\RIGHT{r}
\def\tightcdots{\mathinner{\cdotp\!\cdotp\!\cdotp\mskip -0.5\thinmuskip}}

\begin{document}
\maketitle

\begin{abstract}
  Consider laying out a fixed-topology binary tree of $N$ nodes
  into external memory with block size $B$
  so as to minimize the worst-case number of block memory transfers
  required to traverse a path from the root to a node of depth~$D$.
  We prove that the optimal number of memory transfers is
  $$ \cases{
       \displaystyle
       \Theta\left( {D \over \lg (1{+}B)} \right)
         & when $D = O(\lg N)$, \medskip \cr
       \displaystyle
       \Theta\left( {\lg N \over \lg \left(1{+}{B \lg N \over D}\right)} \right)
         & when $D = \Omega(\lg N)$ and $D = O(B \lg N)$, \medskip \cr
       \displaystyle
       \Theta\left( {D \over B} \right)
         & when $D = \Omega(B \lg N)$.
     } $$

\end{abstract}

\section{Introduction}

Trees can have a meaningful topology in the sense that edges carry
a specific meaning---such as letters from an alphabet in a suffix tree or
trie---and consequently nodes cannot be freely rebalanced.
Large trees do not fit in memory,
so a natural problem is to lay out (store) a tree on disk in a way that minimizes
the cost of a root-to-node traversal.

The \emph{external-memory model} \cite{Aggarwal-Vitter-1988}
(or I/O model or Disk Access Model) defines a memory hierarchy of two levels:
one level is fast but has limited size, $M$, and the other level is slow
but has unlimited size.  Data can be transferred between the two levels
in aligned blocks of size $B$, and an algorithm's performance in this model
is the number of such \emph{memory transfers}.
An external-memory algorithm may be parameterized by $B$ and~$M$.

The general objective in a \emph{tree-layout problem} is to store the $N$ nodes
of a static fixed-topology tree in a linear array so as to minimize the number
of memory transfers incurred by visiting the nodes in order along a path,
starting at the root of the tree and starting from an empty cache.
The specific goal in a tree-layout problem varies depending on the relative
importance of the memory-transfer cost of different root-to-node paths.
(It is impossible to minimize the number of memory transfers along
every root-to-node path simultaneously.)

Tree-layout problems have been considered before.
Clark and Munro~\cite{Clark-Munro-1996} give a linear-time algorithm
to find an external-memory tree layout
with the minimum worst-case number of memory transfers
along all root-to-leaf paths.
Gil and Itai~\cite{Gil-Itai-1999} give a polynomial-time algorithm
to find an external-memory tree layout
with the minimum expected number of memory transfers
along a randomly selected root-to-leaf path, given a fixed independent
probability distribution on the leaves.

\paragraph{Our results.}
We consider the natural parameterization of the tree-layout problem
by the \emph{length} $D$ of the root-to-node path, i.e., the maximum depth $D$
of the accessed nodes.  
We characterize the worst-case number of memory transfers incurred by a
root-to-node path in a binary tree,
over all possible values of these parameters, as
  $$ \cases{
       \displaystyle
       \Theta\left( {D \over \lg (1{+}B)} \right)
         & when $D = O(\lg N)$, \medskip \cr
       \displaystyle
       \Theta\left( {\lg N \over \lg \left(1{+}{B \lg N \over D}\right)} \right)
         & when $D = \Omega(\lg N)$ and $D = O(B \lg N)$, \medskip \cr
       \displaystyle
       \Theta\left( {D \over B} \right)
         & when $D = \Omega(B \lg N)$.
     } $$

This characterization consists of an external-memory 
layout algorithm, and a matching worst-case lower bound.
In particular we show that the optimal cost does not depend on the cache
size~$M$:
our layout assumes a cache just big enough to store a single block ($M=B$),
while the lower bound applies to an arbitrary cache (provided each search
operation starts with an empty cache).
The external-memory layout algorithm runs in $O(N)$ time;
the same upper bound trivially also holds on the number of memory transfers.
As in previous work, we do not know how to guarantee a substantially smaller
number of memory transfers during construction, because on input the tree
might be scattered throughout memory.

\section{Upper Bound}

Our layout algorithm consists of two phases.  The first phase is simple
and achieves the desired bound for $D = O(\lg N)$ without significantly
raising the cost for larger~$D$.  The second phase is more complicated,
particularly in the analysis, and achieves the desired bound for
$D = \Omega(\lg N)$.
Both phases run in $O(N)$ time.

\subsection{Phase 1}

The first part of our layout simply stores the first $\Theta(\lg N)$ levels
according to a B-tree clustering, as if those levels contained a perfect binary
tree.
More precisely, the first block in the layout consists of the $\leq B$
nodes in the topmost $\lfloor \lg (B+1) \rfloor$ levels of the binary tree.
Conceptually removing these nodes from the tree leaves $O(B)$ disjoint trees
which we lay out recursively, stopping once the topmost $c \lg N$ levels
have been laid out, for any fixed $c > 0$. All the data in this layout is stored contiguously; no extra space is left should the top levels not form a complete tree.

This phase defines a layout for a subtree of the tree, which we call the
\emph{phase-1 tree}.  The remaining nodes form a forest of nodes to be
laid out in the second phase.  We call each tree of this forest
a \emph{phase-2 tree}.

The number of memory blocks along any root-to-node path within the phase-1
tree, i.e., of length $D \leq c \lg N$,
is $\Theta(D / \lg (B+1))$.
More generally, any root-to-node path incurs a cost of
$\Theta(\min\{D,\lg N\} / \lg (B+1))$ within the phase-1 tree,
i.e., for the first $c \lg N$ nodes.

\subsection{Phase 2: Layout Algorithm}

The second phase defines a layout for each phase-2 tree, i.e., for each
tree of nodes not laid out during the first phase.

For a node $x$ in the tree,
let $w(x)$ be the \emph{weight} of $x$, i.e., the number of nodes in the
subtree rooted at node~$x$.
Let $\LEFT(x)$ and $\RIGHT(x)$ be the left and right children of node~$x$,
respectively.
If $x$ lacks a child, $\LEFT(x)$ or $\RIGHT(x)$ is a null node
whose weight is defined to be~$0$.

For a simpler recursion, we consider a generalized form of the layout problem
where the goal is to lay out the subtree rooted at a node $x$ into blocks
such that the block containing the root of the tree is constrained to have
at most $A$ nodes, for some nonnegative integer $A \leq B$,
while all other blocks can have up to $B$ nodes.
This restriction represents the situation when $B-A$ nodes have already been
placed in the root block (in the caller to the recursion),
so space for only $A$ nodes remains.

Our algorithm chooses a set $K(x,A)$ of nodes to store in the root block
by placing the root $x$ and dividing the remaining $A-1$ nodes
of space among the two children subtrees of $x$ proportionally
according to weight.
More precisely, $K(x,A)$ is defined recursively as follows:
$$K(x,A) =
    \cases{
      \emptyset & if $A < 1$, \cr
      \{ x \} \cup K[\LEFT(x), (A-1) \cdot w(\LEFT(x))/w(x)] \cr
      \setbox1\hbox{$\{x\}$}
      \hspace*{\wd1}{}
              \cup K[\RIGHT(x), (A-1) \cdot w(\RIGHT(x))/w(x)]
                & otherwise.
    }$$
Because $w(x) = 1 + w(\LEFT(x)) + w(\RIGHT(x))$,
$|K(x,A)| \leq A$.
Also, for $A \geq 1$, $K(x,A)$ always includes the root node $x$ itself.

At the top level of recursion, the algorithm creates a root block $K(r,B)$,
where $r$ is the root of the phase-2 tree $T$, as the first block in the layout
of that tree~$T$.
Then the algorithm recursively lays out the trees in the forest
$T - K(r,B)$, starting with root blocks of $K(r',B)$ for each child $r'$
of a node in $K(r,B)$ that is not in $K(r,B)$.

\subsection{Phase 2: Analysis}
\label{phase 2 analysis}

Within this analysis, let $D_2$ denote the depth of the path within the
phase-2 tree $T$ of the path under consideration ($\Theta(\lg N)$ less than the global notion of $D$).
Define the \emph{density} $d(x)$ of a node $x$ to be
$w(x)/w(r)$ where $r$ is the root of the phase-2 tree $T$.
In other words, the density of $x$ measures the fraction of the entire tree
within the subtree rooted at the node~$x$.
Let $T_x$ denote the subtree rooted at~$x$.

Consider a (downward) root-to-node path $x_0, x_1, \ldots, x_k$
where $x_0$ is the root of the tree.
Define $d_i = d(x_i)$ for $0 \leq i \leq k$,
and define $q_i = d_i/d_{i-1}$ for $1 \leq i \leq k$.
Thus $d_i = d_0 q_1 q_2 \cdots q_i = q_1 q_2 \cdots q_i$ because $d_0 = 1$.
If $x_k$ is in the block containing the root~$x_0$, then
the number $m_k$ of nodes from $T_{x_k}$ that the algorithm places into that
block is given by the recurrence
\begin{eqnarray*}
  m_0 &=& B \\
  m_k &=& (m_{k-1} - 1) q_k
\end{eqnarray*}
which solves to
\begin{eqnarray*}
  m_k
  &=& \underbrace{((\tightcdots(((}_{k}
        B-1) q_1 - 1) q_2 - 1) q_3 \cdots - 1) q_{k-1} - 1) q_k \\
  &=& (B q_1 q_2 \cdots q_k) - (q_1 q_2 \cdots q_k) - (q_2 q_3 \cdots q_k)
    - \cdots
    - (q_{k-1} q_k) - (q_k) \\[\medskipamount]
  &=& B d_k - d_k - {d_k \over d_1} - \cdots
    - {d_k \over d_{k-2}} - {d_k \over d_{k-1}}.
\end{eqnarray*}
This number is at least~$1$ precisely when there is room for $x_k$
in the block containing the root~$x_0$.
Thus, if $x_k$ is not in the block containing the root~$x_0$,
then we must have the opposite:
$$
      B d_k - d_k - {d_k \over d_1} - \cdots
    - {d_k \over d_{k-2}} - {d_k \over d_{k-1}}
    < 1,
$$
i.e.,
$$
      d_k + {d_k \over d_1} + \cdots
    + {d_k \over d_{k-2}} + {d_k \over d_{k-1}}
    > B d_k - 1.
$$
Because $d_0 \geq d_1 \geq \cdots \geq d_k$,
each term $d_k/d_i$ on the left-hand side is at most~$1$,
so the left-hand side is at most~$k$.
Therefore $k > B d_k - 1$.

Let $\f_B(N,D_2) = \f(N,D_2)$ denote the number of memory blocks of size $B$
visited along a worst-case root-to-node path of length $D_2$ in a tree of $N$
nodes laid out according to our algorithm.
Certainly $\f(N,D_2)$ is nondecreasing in $N$ and~$D_2$.
Suppose the root-to-node path visits nodes in the order
$x_0, x_1, \dots, x_k, \dots$, with $x_k$ being the first node outside
the block containing the root node.
By the analysis above,
\begin{eqnarray*}
  \f(N,D_2) & =  & \f(N d_k, D_2 - k) + 1 \\
          &\leq& \f(N d_k, D_2 - d_k B + 1 ) + 1.
\end{eqnarray*}

This inequality is a recurrence that provides an upper bound on $\f(N,D_2)$.
The base cases are $\f(1,D_2) = 1$ and $\f(N,0) = 1$.
In the remainder of this section, we solve this recurrence.

%

Define $x_{k_0}, x_{k_1}, x_{k_2}, \dots, x_{k_t}$ to be the first node within
each memory block visited along the root-to-node path.  Thus, $x_{k_j}$ is the
root of the subtree formed by the $j$th block, so $x_{k_0}$ is the root of
the tree, and $k_1 = k$.
As before, define $d_{k_j} = d(x_{k_j})$.
Now we can expand the recurrence $t$ times:
$$\f(N,D_2) \leq \f\left(N \prod_{i=1}^t d_{k_i},
                       D_2 - B \sum_{i=1}^t d_{k_i}+t\right) + t.$$

So the $\f(N,D_2)$ recursion terminates when
$$\prod_{i=1}^t d_{k_i} \leq {1 \over N}
  \quad \mbox{or} \quad
  \sum_{i=1}^t d_{k_i} \geq {D_2 + t \over B},$$
whichever comes first.
Because $t \leq D_2$, the recursion must terminate once
$$\prod_{i=1}^t d_{k_i} \leq {1 \over N}
  \quad \mbox{or} \quad
  \sum_{i=1}^t d_{k_i} \geq {2 D_2 \over B},$$
whichever comes first.

Our goal is to find an upper bound on the maximum value of $t$ at which the
recursion could terminate, because $t+1$ is the number of memory transfers
incurred.  Define $p$ to be the average of the $d_{k_i}$'s,
$(d_{k_1} + \cdots + d_{k_t})/t$.
In the termination condition,
the product $\prod_{i=1}^t d_{k_i}$ is at most $\prod_{i=1}^t p$
because the product of terms with a fixed sum is maximized when the terms
are equal;
and the sum $\sum_{i=1}^t d_{k_i}$ is equal to $\sum_{i=1}^t p$.
Thus, the following termination condition is satisfied no earlier than the
original termination condition:
$$\prod_{i=1}^t p \leq {1 \over N}
  \quad \mbox{or} \quad
  \sum_{i=1}^t p \geq {2 D_2 \over B}.$$
Therefore, by obtaining a worst-case upper bound on $t$ with this termination
condition, we also obtain a worst-case upper bound on~$t$ with the original
termination condition.

Now
the $\f(N,D_2)$ recursion terminates when
$$p^t \leq  {1 \over N}
  \quad \mbox{or} \quad
  t p \geq {2 D_2 \over B},$$
i.e., when
$$t \geq {\lg N \over \lg (1/p)}
  \quad \mbox{or} \quad
  t \geq {2 D_2 \over B p},$$

Thus we obtain the following upper bound on the number of memory transfers
along this path:
$$t+1 \leq \min\left\{{\lg N \over \lg (1/p)}, {2 D_2 \over B p}\right\} + 2.$$

Maximizing this bound with respect to $p$ gives us an upper bound irrespective
of~$p$.  The maximum value is achieved in the limit when either $p=0$, $p=1$, or
the two terms in the $\min$ are equal.  As $p \to 0$, the bound converges to $0$,
so this is never the maximum.  As $p \to 1$, the bound converges to $2 D_2/B$.
The two terms in the $\min$ are equal when, by cross-multiplying,
\begin{equation} \label{p eqn}
  B p \lg N = 2 D_2 \lg (1/p),
\end{equation}
i.e.,
$$ {1 \over p} \lg {1 \over p} = {B \lg N \over 2 D_2}, $$
or asymptotically
\begin{equation} \label{p soln}
  {1 \over p} = \Theta\left( {{B \lg N \over D_2} \over
                              \lg \left(2{+}{B \lg N \over D_2}\right)} \right).
\end{equation}
In this case, the $\min$ terms are
$$ \Theta\left( {\lg N \over \lg \left(2{+}{B \lg N \over D_2}\right)} \right).
$$

Therefore, the upper bound is
$$ \max\left\{O\left({\lg N \over \lg \left(2{+}{B \lg N \over D_2}\right)}
               \right),
              {D_2 \over B}\right\},
$$
or
$$ O\left({\lg N \over \lg \left(2{+}{B \lg N \over D_2}\right)}
        + {D_2 \over B}\right). $$

\subsection{Putting It Together}

The total number of memory transfers is the sum over the first and second
phases.  If $D \leq c \lg N$, only the first phase plays a role, and the
cost is $O(D / \lg (B+1))$.  If $D > c \lg N$, the cost is the sum
$$ O\left(
     {c \lg N \over \lg (B+1)} +
     {\lg N \over \lg \left(2{+}{B \lg N \over D - c \lg N}\right)} +
     {D - c \lg N \over B}
   \right), $$
which is at most
$$ O\left(
     {c \lg N \over \lg (B+1)} +
     {\lg N \over \lg \left(2{+}{B \lg N \over D}\right)} +
     {D \over B}
   \right). $$
Because $D = \Omega(\lg N)$, the denominator of the second term is at most
$\lg (B+1)$, so the first term is always at most the second term up to constant
factors.  Thus we focus on the second and third terms.
If $D = X \lg N$, then the second term is $O((\lg N) / \lg (2 + B/X))$
and the third term is $O((X \lg N) / B) = O((\lg N) / (B/X))$.
For $X = O(B)$, the second term divides $\lg N$ by $\Theta(\lg (B/X))$,
while the third term divides $\lg N$ by $\Theta(B/X)$.
Thus the second term is larger up to constant factors for $X = O(B)$.
For $X = \Omega(B)$, the second term is $O(\lg N)$,
while the third term is $O((X/B) \lg N)$, which is larger up to
constant factors.

In summary, the first term dominates when $D = O(\lg N)$,
the second term dominates when $D = \Omega(\lg N)$ and $D = O(B \lg N)$,
and the third term dominates when $D = \Omega(B \lg N)$.
Therefore we obtain the following overall bound:

\begin{theorem} \label{upper bound}
  Given $B$ and a fixed-topology tree on $N$ nodes, we can compute in $O(N)$
  time an external-memory tree layout with block size $B$
  in which the number of memory transfers
  incurred along a root-to-node path of length~$D$ is
$$ O\left(
     \cases{
       \displaystyle
       {D \over \lg (1{+}B)}             & when $D = O(\lg N)$ \medskip \cr
       \displaystyle
       {\lg N \over \lg \left(1{+}{B \lg N \over D}\right)}
                                         & when $D = \Omega(\lg N)$
                                           and $D = O(B \lg N)$ \medskip \cr
       \displaystyle
       {D \over B}                       & when $D = \Omega(B \lg N)$
     } \right). $$
\end{theorem}

\section{Lower Bound}

\subsection{Convexity}

First we describe a useful structural property that can be assumed without
loss of generality of the worst-case optimal tree layout.
A layout is \emph{convex} if every block contains a (possibly empty)
contiguous subpath of any root-to-node path in the tree.
Any convex layout is insensitive to the cache size~$M$
(assuming $M \geq B$), because once a root-to-node path leaves
a block, it never returns to that block; thus, the memory-transfer cost
equals the number of distinct blocks along the path.

We prove that there exists a convex worst-case optimal tree layout.
Our proof mimics the analogous result for minimizing the
expected cost of a root-to-node path \cite[Lemma~3.1]{Gil-Itai-1999}.

First we need some terminology.
Define the \emph{contiguity} of a node $x$ to be the number of nodes in the
tree that are reachable from $x$ while remaining within the block containing~$x$
(i.e., the size of $x$'s connected component in the subgraph
induced by the block containing~$x$).
Define the \emph{contiguity signature} of a tree layout to be the sequence of
contiguities of the nodes of the tree in a consistent order that visits
ancestors before descendants
(e.g., pre-order or left-to-right breadth-first search).

We claim that the worst-case optimal tree layout with the lexically maximum 
contiguity signature is convex.
For any root-to-node path $x_1, x_2, \dots, x_n$, suppose to the contrary
that $x_i$ and $x_j$ ($i < j$) are stored in the same block,
but $x_{i+1}$ is not.
Then we modify the layout by swapping $x_{i+1}$ and $x_j$ (moving $x_{i+1}$ to
the same block as $x_i$, and moving $x_j$ to the block previously
containing~$x_{i+1}$).  This modification can only reduce the set of distinct
blocks visited by any root-to-node path, as $x_{i+1}$ now always gets visited
for free after $x_i$, and any root-to-node path visiting $x_j$ also visits
$x_{i+1}$ so visits the same set of blocks as before the swap.
Therefore the layout remains optimal,
while changing the contiguity vector in two ways.
First, we increment the contiguity of~$x_i$ (and some other nodes).
Second, we decrement the contiguity of $x_j$ and reachable nodes
in the same block, so they must all be descendants of $x_{i+1}$.
Thus the new layout has a lexically larger contiguity signature,
contradicting maximality.

\subsection{Construction}

Now we proceed to the lower-bound construction.
For $D \leq \lg (N+1)$, the perfectly balanced binary tree on $N$ nodes gives a
worst-case lower bound of $\Omega(D/\lg B)$ memory transfers
\cite[Theorem~7]{Nodine-Goodrich-Vitter-1996}.
For all $D$, any root-to-node path of length $D$ requires at least $D/B$
memory transfers just to read the $D$ nodes along the path.
Thus we are left with proving a lower bound for the case
when $D = \Omega(\lg N)$ and $D = O(B \lg N)$.

\begin{figure}[t]
  \centering
  \ifpdf
    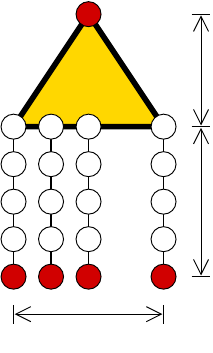
  \else
    \input{lowerbound.pstex_t}
  \fi
  \caption{The recursive lower-bound construction: a complete binary tree
    with $1/p$ leaves attached to $1/p$ paths of length $p B$,
    each attached to a recursive construction.}
  \label{lowerbound}
\end{figure}

The following lower-bound construction essentially mimics the worst-case behavior
predicted in Section \ref{phase 2 analysis}.  We choose $p$ to be the solution
to Equation~\ref{p eqn}, i.e., so that it satisfies $B p \lg N = D \lg (1/p)$.
Because $D = \Omega(\lg N)$, this equation implies that
\begin{equation} \label{relation}
  B p = \Omega(\lg (1/p)).
\end{equation}
The asymptotic solution for $1/p$ is given by Equation~\ref{p soln}:
$$ {1 \over p} = \Theta\left( {{B \lg N \over D} \over
                              \lg \left(2{+}{B \lg N \over D}\right)} \right).
$$
Using this value of $p$, we build a tree of slightly more than $B$ nodes,
as shown in Figure \ref{lowerbound},
that partitions the space of nodes into $1/p$ fractions of~$p$.
We repeat this tree construction recursively in each of the children subtrees,
stopping at the height that results in $N$ nodes.

Consider any convex external-memory layout of the tree.
Because each tree construction has more than $B$ nodes,
it cannot fit in a block.  Thus, every tree construction has at least one
node that is not in the same block as the root.
By convexity, for any $k \leq \log_B N$, there is a root-to-node path that
incurs at least $k$ memory transfers by visiting $k$ distinct blocks
in $k$ tree constructions.
Such a path has length $D = O(k \, [p B + \lg (1/p)])$,
which is $O(k p B)$, by Equation~\ref{relation}.
Therefore
$$k = \Omega\left({D \over p B}\right)
    = \Omega\left({\lg N \over \lg \left(2{+}{B \lg N \over D}\right)}\right).
$$

\begin{theorem}\label{mainth}
  For any values of $N$, $B$, and~$D$, there is a fixed-topology tree on $N$
  nodes in which every external-memory layout with block size $B$ incurs
  $$ \Omega\left(
       \cases{
         \displaystyle
         {D \over \lg (1{+}B)}             & when $D = O(\lg N)$ \medskip \cr
         \displaystyle
         {\lg N \over \lg \left(1{+}{B \lg N \over D}\right)}
                                           & when $D = \Omega(\lg N)$
                                             and $D = O(B \lg N)$ \medskip \cr
         \displaystyle
         {D \over B}                       & when $D = \Omega(B \lg N)$
       } \right) $$
  memory transfers along some root-to-node path of length~$D$.
\end{theorem}

\section{Alternate Models}

There are several possible variations on the model considered here.
We assume that every traversal follows a root-to-leaf path, following
child pointers from one node to the next.  In this model, it does not
make sense to store a node in more than one block, because there is only
one way to reach each node, so only one copy could ever be visited.
However, if we allow multiple versions of a pointer that lead to different
copies of a node, we could imagine doing better---indeed, with unlimited space,
we can easily achieve $O(D/B)$ search cost by storing a different tree for
every possible leaf.  An interesting open problem would be to characterize
the trade-off between space and search cost.

The \emph{String B-Tree} data structure~\cite{DBLP:journals/jacm/FerraginaG99}
also seeks to support efficient tree operations in external memory for the
purpose of implementing various string operations. The performance of their
structure is identical to our bounds as stated in Theorem~\ref{mainth} in the
two extreme ranges, but outperforms ours slightly in the middle range.
This difference comes from a further difference in model: the string B-tree
effectively stores pointers from nodes to deep descendants, not just
children, allowing a traversal to effectively skip some nodes along the
root-to-node path.  Our results show that such a change in model
is necessary to achieve their runtime.

\section*{Acknowledgments}

This research was initiated during the Seminar on Cache-Oblivious and
Cache-Aware Algorithms held at Schloss Dagstuhl in July 2004.
We thank Jeff Erickson and J. Ian Munro for many helpful discussions
during that seminar.
We also thank Paulo Ferragina for early discussions on this problem.

\bibliography{cache,succinct}
\bibliographystyle{alpha}

\end{document}